\documentclass[journal,10p]{IEEEtran}

\usepackage{cite}
\usepackage[hyphens]{url}
\usepackage[utf8]{inputenc}
\usepackage{graphicx}
\usepackage{float}
\usepackage{color, colortbl}
\usepackage{xcolor}
\usepackage{array}
\usepackage{multirow}
\usepackage{footnote}
\usepackage[normal]{threeparttable}
\usepackage{mathtools}

\usepackage{adjustbox}
\usepackage{subfig}
\usepackage{siunitx}

\makeatletter
\newcommand{\removelatexerror}{\let\@latex@error\@gobble}
\makeatother

\makeatletter

\newcommand{\Rmnum}[1]{\expandafter\@slowromancap\romannumeral #1@}
\makeatother
\usepackage{graphicx,epsfig}
\usepackage{amsmath,amssymb,amsfonts,nccmath}
\usepackage{mathrsfs}
\usepackage{algorithm}
\usepackage{algorithmicx}
\usepackage{algpseudocode}
\usepackage{ulem}
\usepackage{diagbox}
\usepackage{authblk}
\usepackage{comment}
\usepackage{cleveref}
\newcommand*{\depaddr}[1]{\dagmark Computer and Information Sciences, \asmark Electrical and Computer Engineering} 

\newcommand*{\instaddr}[1]{\dagmark University of Delaware~~~~~ \ddagmark Cisco Systems}

\newcommand*{\dagmark}[1][\dag]{\textsuperscript{\dag}}
\newcommand*{\asmark}[1][*]{\textsuperscript{*}}
\newcommand*{\ddagmark}[1][\ddag]{\textsuperscript{\ddag}}
\newcommand*{\email}[1]{\texttt{#1}}

\begin{document}

\title{Traffic Characteristics of Virtual Reality  \\ over Edge-enabled Wi-Fi Networks}

\author{%
Seyedmohammad Salehi\dagmark[1]~~~ Abdullah Alnajim\dagmark[1]~~~
Xiaoqing Zhu\ddagmark[1]~~~ Malcolm Smith\ddagmark[1] ~~~~~~~~~~~~~~~~~~~~~~~~~~~~~~~Chien-Chung Shen\dagmark[1]~~~ Leonard Cimini\dagmark[1]  \\
\instaddr{\dagmark[1]}  \\
\email{\{salehi,alnajim,cshen,cimini\}@udel.edu, \{xiaoqzhu,mmsmith\}@cisco.com}
}

\maketitle

\begin{abstract}
Virtual reality (VR) is becoming prevalent with a plethora of applications in education, healthcare, entertainment, etc. To increase the user mobility, and to reduce the energy consumption and production cost of VR head mounted displays (HMDs), wireless VR with edge-computing has been the focus of both industry and academia. However, transferring large video frames of VR applications with their stringent Quality of Service (QoS) requirements over wireless network requires innovations and optimizations across different network layers. In order to develop efficient architectures, protocols and scheduling mechanisms, the traffic characteristics of various types of VR applications are required. In this paper, we first compute the theoretical throughput requirements of an ideal VR experience as well as a popular VR HMD. We then examine the traffic characteristics of a set of VR applications using an edge-enabled Wi-Fi network. Our results reveal interesting findings that can be considered in developing new optimizations, protocols, access mechanisms and scheduling algorithms. 

\end{abstract}


\section{Introduction}
\label{sec-intro}

Virtual reality (VR) is a technology with myriads of use cases in education, healthcare, entertainment, gaming, to name a few. VR applications are sensitive to latency and packet loss, and thus, require reliable and low-latency access to powerful graphics processors to render the projected high-resolution scenes. Due to the limited graphics processing capability of the VR head-mounted displays (HMDs) and to reduce the energy consumption of HMDs for graphics processing and the HMD's production cost, for a satisfying immersive experience, VR HMDs are typically connected, via a cable, to a server with a powerful graphics card for video frame\footnote{Application layer packet data unit (APDU) with video content.} rendering. In this architecture, although the HMD can access a server with advanced graphics capabilities, it becomes mobility limited. 

Newer HMDs (e.g., Oculus Quest) are equipped with not only more capable GPU modules to independently render the video contents, but also wireless capability to facilitate untethered mobility. Although these HMDs allow more mobility, the on-device rendering consumes more energy and cannot keep up with the emerging very high-resolution contents. Nonetheless, since these HMDs are equipped with wireless modules, most of the rendering can be offloaded to an edge server in close proximity. Compared to cloud computing, edge/fog computing \cite{bonomi2014fog,chen2017fog,hu2015mobile} brings the computation capability closer to the end user, which not only reduces the amount of traffic transferred over the Internet but also reduces the end-to-end latency. This architecture, however, incurs a heavy traffic load on the wireless access network: large, delay-sensitive VR physical layer protocol data units (PPDUs) with high video frame rate in the downlink (DL) and high frequency user's tracking information (TI) in the uplink (UL), whereby most existing Wi-Fi standards (e.g., IEEE 802.11n and 802.11ac) are unable to guarantee the desired Quality of Service (QoS) requirements, especially in dense environments. 

To meet the current and emerging traffic demands, IEEE 802.11 has evolved to 802.11ax \cite{bellalta2016ieee,khorov2018tutorial} (employing OFDMA, beamforming, multiuser (MU) MIMO, and more efficient access mechanisms) and 802.11ad/ay (employing millimeter-wave) \cite{sun2017novel}. However, various VR applications have different traffic load, delay and bandwidth requirements so that there exists a trade-off between satisfying the QoS requirements of VR applications and the overall throughput of the wireless networks. Thus, it is important to understand the traffic characteristics of different VR applications in order to design architectures, protocols and access mechanisms for Wi-Fi networks that satisfy the QoS requirements of specific VR applications even in dense deployments. 


\begin{figure}[!htpb]
    \centering 
    \includegraphics[width=\linewidth]{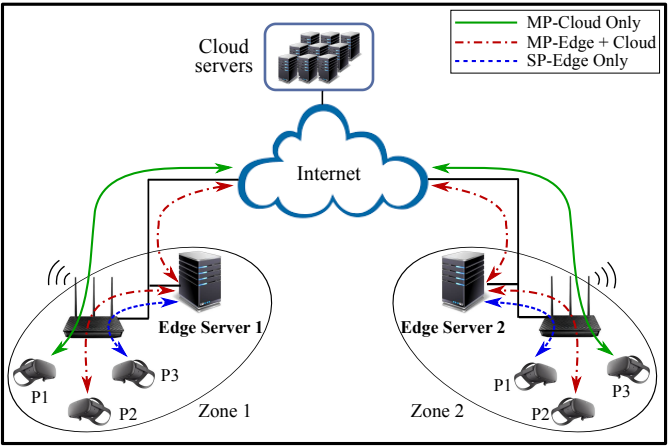} 
    \caption{VR communications with edge and cloud servers. Frame rendering can take place on the HMDs or on the edge servers. In SP-VR applications, no communications with cloud is necessary. However, in MP-VR applications, edge servers/HMDs send/receive shared user information (e.g., voice). For instance, user P2 in Zone 1 communicates with user P3 in Zone 2, although their frame rendering mode may be different.} 
    \label{fig:arch}
\end{figure}

At a high level, VR applications can be classified into single-party (SP) and multi-party (MP). Intuitively, in an SP-VR application where HMD renders the frames, no communication with cloud or edge is necessary. Fig. \ref{fig:arch} depicts the communications between HMD, edge and cloud for SP-VR and MP-VR applications. Frame rendering can either take place on the HMDs or at the edge servers. However, unlike SP-VR experience, in MP-VR, each party also communicates with a cloud server (either directly or via an edge server) to transfer and receive the shared user information (e.g., voice, updated avatar positions, etc.). It is also possible for the edge servers to communicate directly with each other in a peer-to-peer manner. For instance, in cooperative VR games, multiple players share noticeable amount of background in each scene.
In such situations, the background layer can be rendered, cached and reused at the edge servers and the ``delta" images rendered in real-time on the HMD \cite{li2018muvr}. Similarly, Furion \cite{lai2019furion} separated the image rendering between the edge server and the mobile device for Google's Daydream VR HMDs. Vivo by Han et al. \cite{han2020vivo} optimized the data transfer by predicting user viewport. EC+ by Zhang et al. \cite{zhang2017towards} investigated the segregation of service placement across HMD, edge and cloud with user mobility.

As the UL transmissions of user's TIs are very frequent, in multi-party applications, wireless channel access contentions are unavoidable. Ahn et al. \cite{ahn2018virtual} proposed to reduce the priority of the TI frames so as to reduce contention in 802.11ax networks. The authors further proposed an algorithm to dynamically adjust the frequency of the 802.11ax trigger frames and the video frame rate based on the predefined thresholds of packet loss ratio. Due to the large physical layer transmission delay of the current Wi-Fi standards, Zhang et al. \cite{zhang2018wifi} employed multiple Wi-Fi network interface cards to reduce the transmission delay and jitter of VR frames. Abari et al. \cite{abari2017enabling} investigated the use of mmWave for a low-latency untethered VR experience.

Since VR gaming can be a paradigmatic example of applications requiring complex 3D rendering, Premsankar et al. \cite{premsankar2018edge} employed an open-source cloud gaming platform, GamingAnywhere \cite{huang2014gaminganywhere}, to compare the response delay of edge computing and cloud computing. To achieve an eye-like VR experience, in \cite{bastug2017toward}, the authors analyzed the required throughput by comparing it to the real eye experience. In this paper, we compute the required throughput of VR applications for an eye-like experience as well as an acceptable video quality experience based on the specifications of one of the popular VR HMDs (i.e., Oculus Quest). Moreover, we present a case study of running different VR applications on an open source remote VR display (i.e., ALVR) and characterizing their traffic statistics. Our results are useful for developing new architectures and scheduling mechanisms for future Wi-Fi standards. 

The rest of this paper is organized as follows. We explain the theoretical aspects of the required VR throughput in the next section. In Section \ref{sec:casestudy}, we describe our testbed, network settings and traffic collection choices. In Section \ref{sec:results}, we present the statistics of the experiments along with discussions. Section \ref{sec:conclusion} concludes this paper with future research directions. 

\section{VR throughput requirements}
\label{sec:theoretical}

\subsection{Preliminaries}
The resolution or quality of an image is measured by the number of pixels that represent it. Since resolution depends on the size of the display, pixel per inch (PPI) can be more representative of the image quality. PPI is defined to be the ratio of the number of pixels along a dimension of a display to the length of that dimension. 

\begin{equation}
\label{eq:ppi}
\text{PPI} = \frac{D_{px}}{D_{in}}
\; \; or \; \;
\frac{W_{px}}{W_{in}}
\; \; or \; \;
\frac{H_{px}}{H_{in}},
\end{equation}
where $D_{px}$, $W_{px}$, and $H_{px}$ denote the number of pixels along the diagonal, the width, and the height of a display, respectively. Likewise, $D_{in}$, $W_{in}$, and $H_{in}$ denote, in inches, the length of the diagonal, the width, and the height of the same display, respectively.

\begin{figure}[!hbpt]
    \centering 
\includegraphics[width=\linewidth]{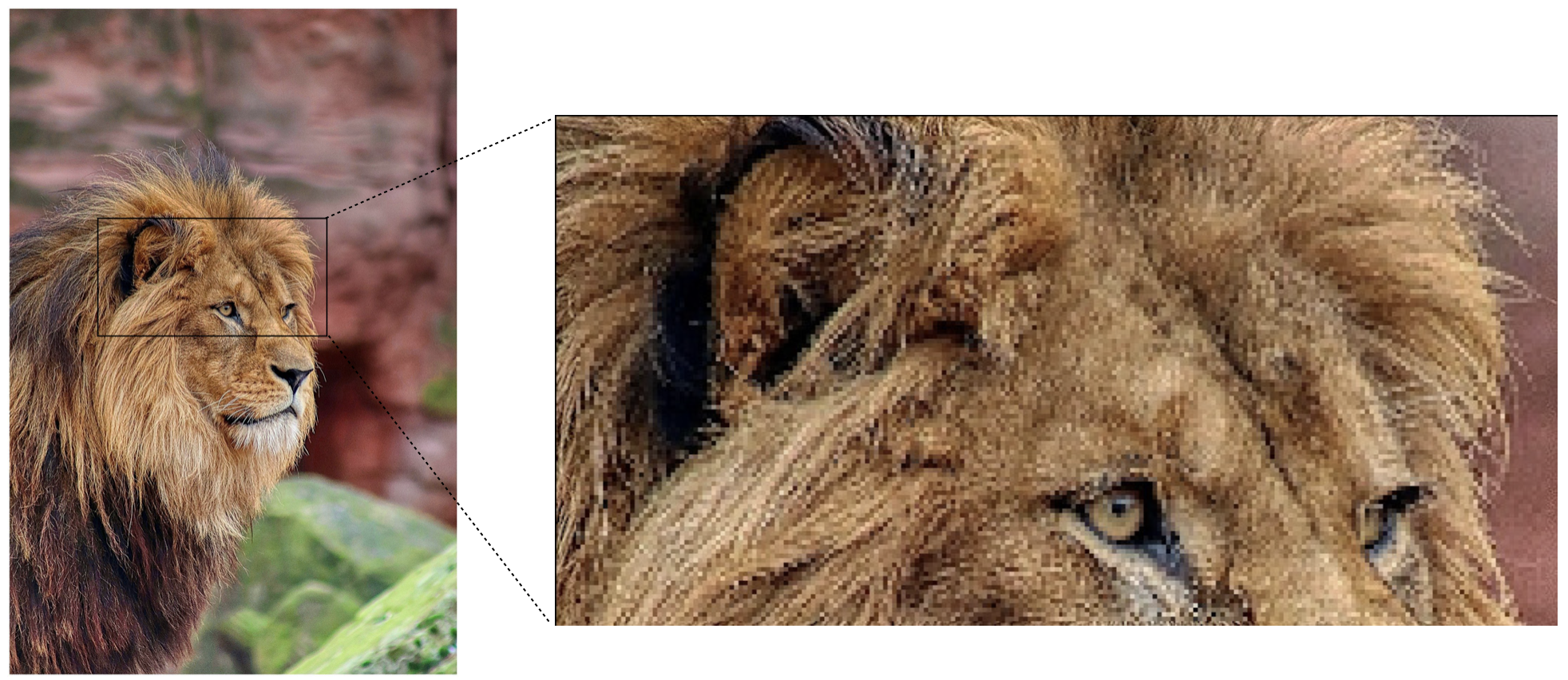} 
    \caption{The effect of distance on the quality of an image. The closer the distance of eyes to an image, the higher resolution is required to observe the same image quality.} 
    \label{fig:distEffect} 
\end{figure}

\begin{figure}[!htpb]
    \centering 
\includegraphics[width=\linewidth]{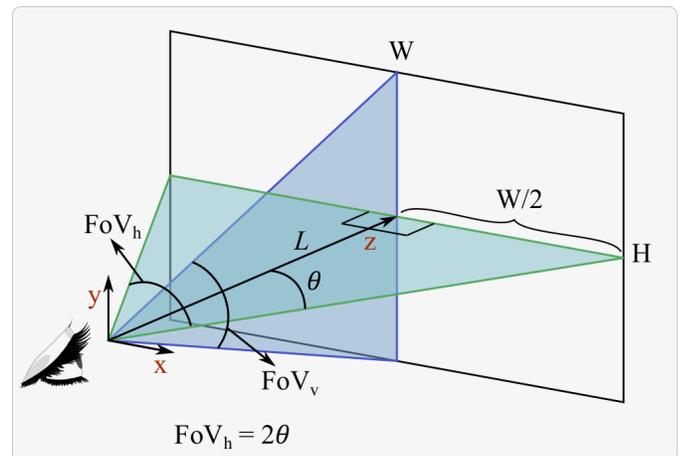} 
    \caption{Computing the Field of View (FoV) using the display width (W) and the viewing distance ($L$). $\theta=\arctan{(\frac{W}{2L})}$. } 
    \label{fig:FoV}
\end{figure}

\begin{figure}[!htpb]
    \centering 
\includegraphics[width=\linewidth]{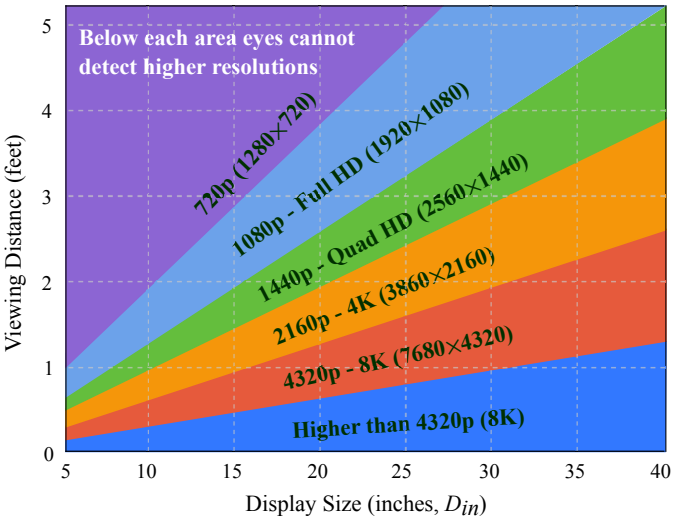}
    \caption{Optimal viewing distance vs. display size. Closer distances require higher display resolution.} 
    \label{fig:disp-dist}
\end{figure}

Image quality also depends on the distance of the display from the eyes as illustrated by Fig. \ref{fig:distEffect}. In that sense, pixels per degree (PPD) is a more accurate measurement of image quality which denotes the number of covered pixels per unit degree from an eye's fovea to the display. As the distance between the screen and the eye fovea decreases, the size of the display area, covered by a unit degree, becomes smaller, so that eyes can detect more details of the displayed image. To compute PPD, we should first compute the Field of View (FoV) which is the maximum amount of the world observable by eyes and depends on the viewing distance and the screen size. For instance, to compute the horizontal FoV ($\rm{FoV}_h$), as illustrated in Fig. \ref{fig:FoV}, we can create two right triangles from the distance of the eye to the screen and, based on the following equation, compute $\rm{FoV}_h$ as 2$\theta$.

\begin{equation}
\label{eq:FoV}
\rm{FoV}_{h} = 2 \times \underbrace{tan^{-1}\bigg(\frac{ 
\rm{W}
}{2 \times L}\bigg)}_{\theta}
\end{equation}
where $\rm{FoV}_{h}$ denotes the horizontal Field of View in degree\footnote{If $\theta$ is in radian, FoV should be converted to degrees by multiplying by \ang{57.3}.}, and $L$ denotes the distance between the user's eyes and the screen in inches. We can now compute $\rm{PPD}_h$ as follows.

\begin{equation}
\label{eq:ppd}
\rm{PPD}_h = \frac{W_{px}}{FoV_{h}}
\end{equation}
%
From the Eq. \ref{eq:ppd}, with the knowledge of $\rm{FoV}_h$ and $\rm{PPD}_h$, we can compute the $\rm{W}_{px}$. Computation of the vertical FoV, $\rm{FoV}_v$, and vertical PPD, $\rm{PPD}_v$, is similar. Fig. \ref{fig:disp-dist} shows the optimal viewing distance for different display sizes with different resolutions. The visual acuity threshold is computed based on 1 arcminute for normal vision (20/20) \cite{wiki_vd_ds}.

To capture the required throughput in this study, we consider two cases: an ideal VR experiment (i.e., an eye-like experiment) and a real VR HMD experience with Oculus Quest. 
A normal vision (20/20) has the binocular horizontal and vertical FoV of \ang{150} and \ang{120}, respectively. The required throughput for an eye-like experience can be computed based on the following equation.

\begin{equation}
\label{eq:eye-throghput}
\rm{T}_{eye} = W_{px} \times H_{px}\times b_{d} \times R_f
\end{equation}
where from Eq. \ref{eq:ppd}, $\rm{W}_{px} = FoV_h \times PPD_h$ and $\rm{H}_{px} = FoV_v \times PPD_v$; $\rm{b}_{d}$ denotes bit depth used to represent a color image
(e.g., to represent three colors each with 256 different intensities, 3$\times$8 bits are required),
and $\rm{R}_f$ denotes the number of frames per second (FPS). 
With a typical PPD of 200 \cite{bastug2017toward},
36 bits for full color representation, and frame rate of 150 FPS, the number of bits that is required to represent this image quality per second is 3.888 Terra bits.

\subsection{Throughput Requirements for Oculus Quest}

Oculus Quest, a VR HMD created by the Oculus VR division of Facebook, has a dual OLED panel with resolution of 1,440$\times$1,600 per eye. Therefore, the required number of bits to represent a scene per second in this HMD for both eyes can be computed as follows.

\begin{equation}
\label{eq:hmd-throghput}
\rm{T}_{HMD} = 2 \times W_{px} \times H_{px}\times b_{d} \times R_f
\end{equation}

Therefore, with 24 bits for each pixel's color, and frame rate of 72 FPS, the number of required bits per second is $2 \times 1440 \times 1600 \times 24 \times 72 \sim 8$ Gb. In the case of offloading the rendering to the edge server, we need compression algorithms to encode a video frame on the edge server, transfer the frame to the HMD over the network and decode it at the HMD. Since there is a trade off between the maximum video compression and the compression time, the encoding techniques can be tuned. Assuming an average compression factor range of [200-300] \cite{bastug2017toward,pavlic975ffmpeg}, the required throughput will be in the range of [26.5-40] Mbps. Moreover, with higher compression factors and rendering optimizations (mentioned in Section \ref{sec-intro}), this required throughput can be reduced.

\begin{figure}[!hbpt]
    \centering 
\includegraphics[width=\linewidth]{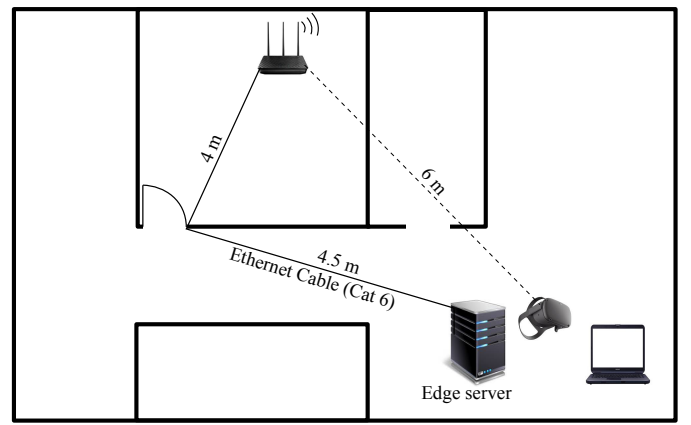} 
    \caption{The topology of our testbed.} 
    \label{fig:topo} 
\end{figure}

\section{Case Study: Oculus Quest}
\label{sec:casestudy}

\subsection{Network Topology and Settings}

Oculus Quest is designed to operate in wireless mode for untethered communication with the cloud servers. Although Oculus Quest can be used in standalone mode for video frame rendering, owing to the wireless capability, frame rendering can also be offloaded to an edge server. To characterize the impact of edge-enabled VR applications over Wi-Fi networks, we deployed a server with a high-end graphics card (AMD Radeon VII) to serve as the edge server. As depicted in Fig. \ref{fig:topo}, this server is connected to a wireless router (ASUS TM-1900AC) via an Ethernet cable. An Oculus Quest HMD is also connected to this router wirelessly using channel 161 of 20 MHz. Oculus Quest can operate on dual bands and support Wi-Fi standards up to IEEE 802.11ac. 

To be able to capture the Wi-Fi frames transmitted between the HMD and the wireless router and decode all the layers of the Internet protocol stack, the wireless network is intentionally made ``open access." Specifically, a laptop is connected to this wireless network to sniff the Wi-Fi frames by using Wireshark. Also, the Ethernet frames between the server and the wireless router are captured using Wireshark running on the server.
In this experiment, we used the open source remote VR display called Air Light VR (ALVR), where the server application runs on the edge server and the client application runs on the HMD. 

\subsection{VR Traffic Collection}

Since VR games can be representatives of complex 3D VR applications, for our analysis, we selected three VR games in ascending required video resolution: \textit{Rec Room}, \textit{The Lab}, and \textit{War Robots}. 
\textit{Rec Room} belongs to the category of multi-player social interactive games, where it allows users to talk, create games and play with each other. \textit{War Robots} and \textit{The Lab} are single-player games. The former is a high-resolution game and the latter offers a series of different game types accessible through a hub room. Specifically, we play \textit{Xortex}, a delay-sensitive shooting game. Traffic traces, source code, and the post-processing scripts can be accessed online \cite{traces}.



\begin{figure}[!htpb]
    \centering 
    \includegraphics[width=\linewidth]{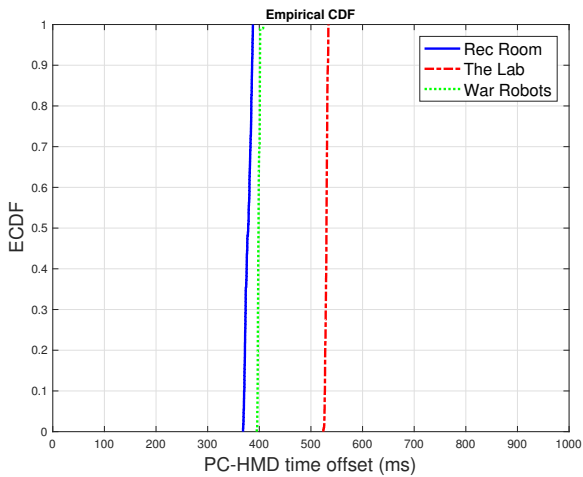} 
    \caption{Time offset between the edge server and the client (HMD).} 
    \label{fig:timeoffset}
\end{figure}

Unlike network simulators where clocks are usually assumed to be synchronized among all the devices, in a client-server application, the accurate computation of the end-to-end latency is subject to the clock offset between the two devices. To get around this problem, one computer can periodically measure\footnote{This measurement can be either from timestamps included in the packet headers or from specific packets to solicit the clock of the other machine.} its clock offset with the second computer and compute the end-to-end latency more accurately. In ALVR, clock offset is measured roughly every 1 second with an especial Application layer packet data unit (APDU) designed to sent back to sender with the timestamp of the receiver.

Fig. \ref{fig:timeoffset} depicts the client-server time offsets in different games. During the course of each game, the time offset does not change significantly, which asserts the accuracy of latency computations at the application layer compared to the PHY/MAC layers.

\begin{figure}[!htbp]
  \includegraphics[width=\linewidth]{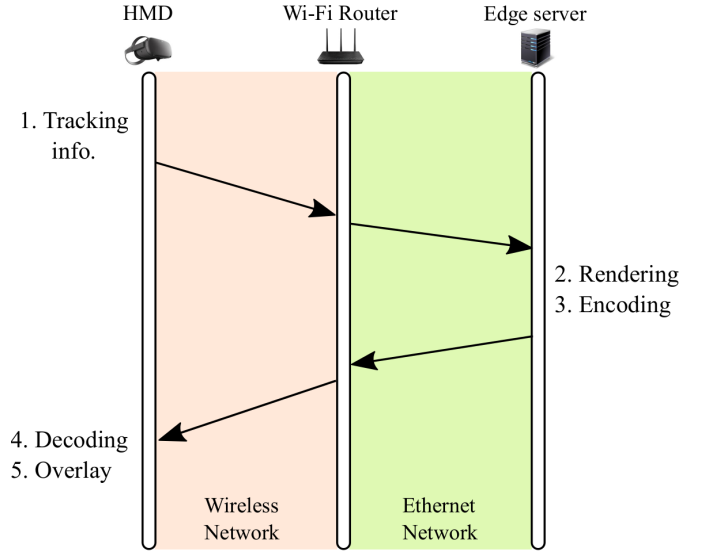}
  \caption{Process of fetching one video frame from the server. }
  \label{fig:timing}
\end{figure}

\subsection{Frame Rendering on the Edge Server}

As illustrated in Fig. \ref{fig:timing}, in a typical VR experience, the HMD (periodically) collects and sends small APDus of tracking information containing the position 
and orientation 
of the user's head and controllers\footnote{A user's hands can also act as controllers.} to the edge server to retrieve the appropriate video frame with size equal to the HMD's 
display resolution. Although such data is usually small, it has a high frequency (e.g., 200-1000 Hz). Consequently, based on the received position and orientation data, the edge server locates, renders, encodes and transmits the target video frame back to the HMD. Such high-resolution video frames will be fragmented into small MAC layer packet data units (MPDUs), traverse over the networks (Ethernet and Wi-Fi) and reassembled at the HMD. After receiving the video frame, the HMD decodes, renders the updated haptics positions on the received video frame as overlay and submits the final rendered video frame to each lens. 

\begin{figure*}[!htbp]
\centering
\includegraphics[width=0.52\linewidth]{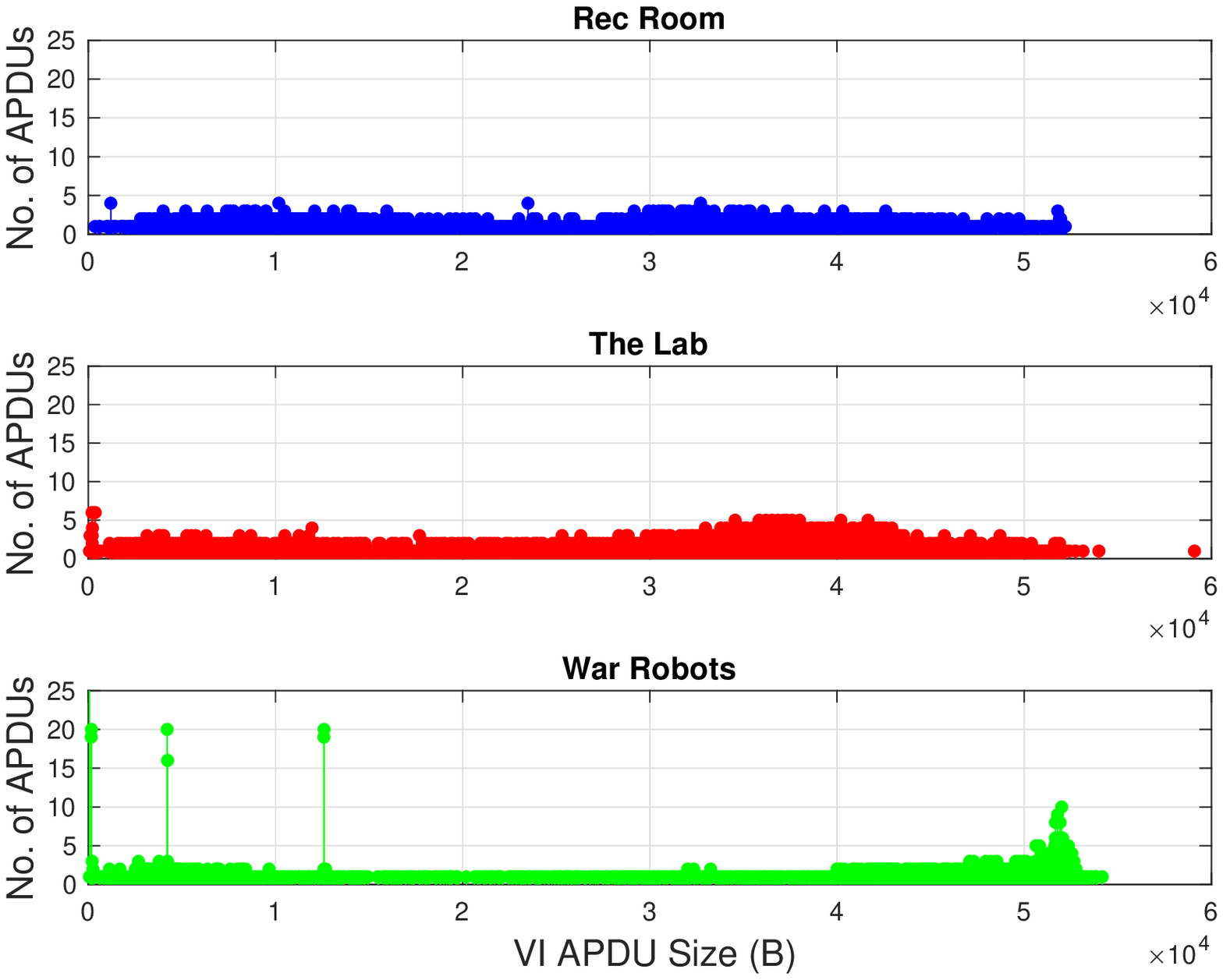} \hspace{-3em}
\includegraphics[width=0.52\linewidth]{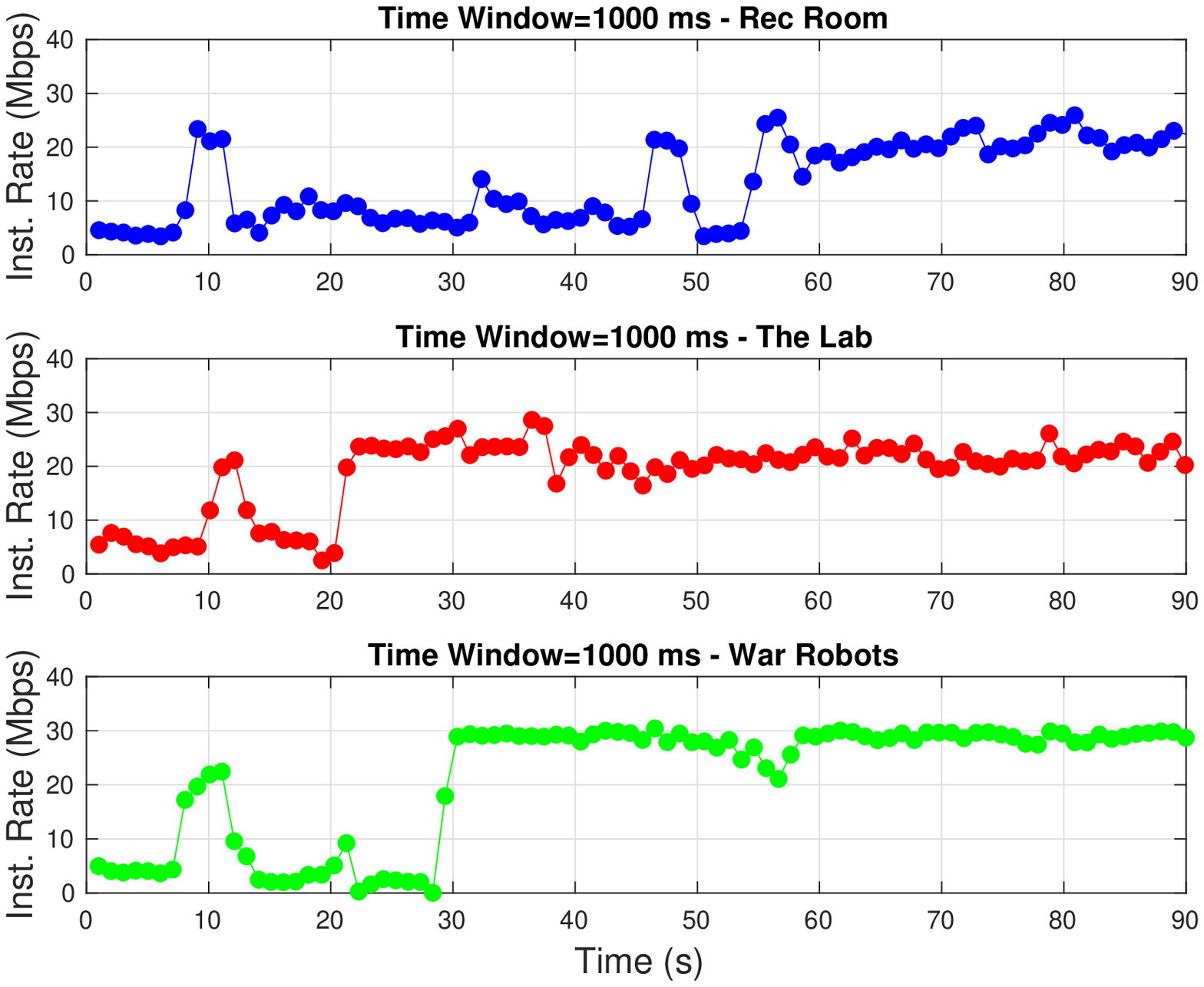}
\caption{Video frame size and the instantaneous rate (time window of 1 s) at the application layer for various VR games. Each experiment lasts for 90 seconds; ALVR client and server connection, load, and play of the game. With higher graphics demand, larger video frames are generated. Average rate the Application layer for Rec Room, The Lab and War Robots are 14.13, 17.61 and 21.42 Mbps, respectively. However, with the overhead of different layers, the average rate at the MAC layer is 17.20, 21.51, and 26.19 Mbps, respectively.}
\label{fig:app_size_num_var}
\end{figure*}

\subsection{H.264 vs. H.265 vs. H.266}
High efficiency video coding (HEVC) or H.265 is originally designed for more efficient transferring of video frames over the wireless networks. H.265 not only has higher compression factor compared to its predecessor, advanced video encoding (AVC) or H.264, it also has better motion compensation and spatial prediction. Therefore, with the same target bitrate and frame rate, H.265 can produce higher quality videos with smaller file sizes compared to H.264 (30-50\% better compression rate). 
In addition, versatile video coding (VVC) or H.266, has recently been standardized specifically for encoding videos of higher resolutions (4K to 16K) as well as \ang{360} videos. With 30-50\% better compression rate compared to HEVC, VVC will enable more efficient transfer of high quality video contents over the Internet \cite{vvc_eric}. Due to unavailability of VVC encoder at the time of this study, we compare the performance of H.264 and H.265 for the game of Rec Room.

\begin{figure*}[!htbp]
\centering
\includegraphics[width=0.495\linewidth]{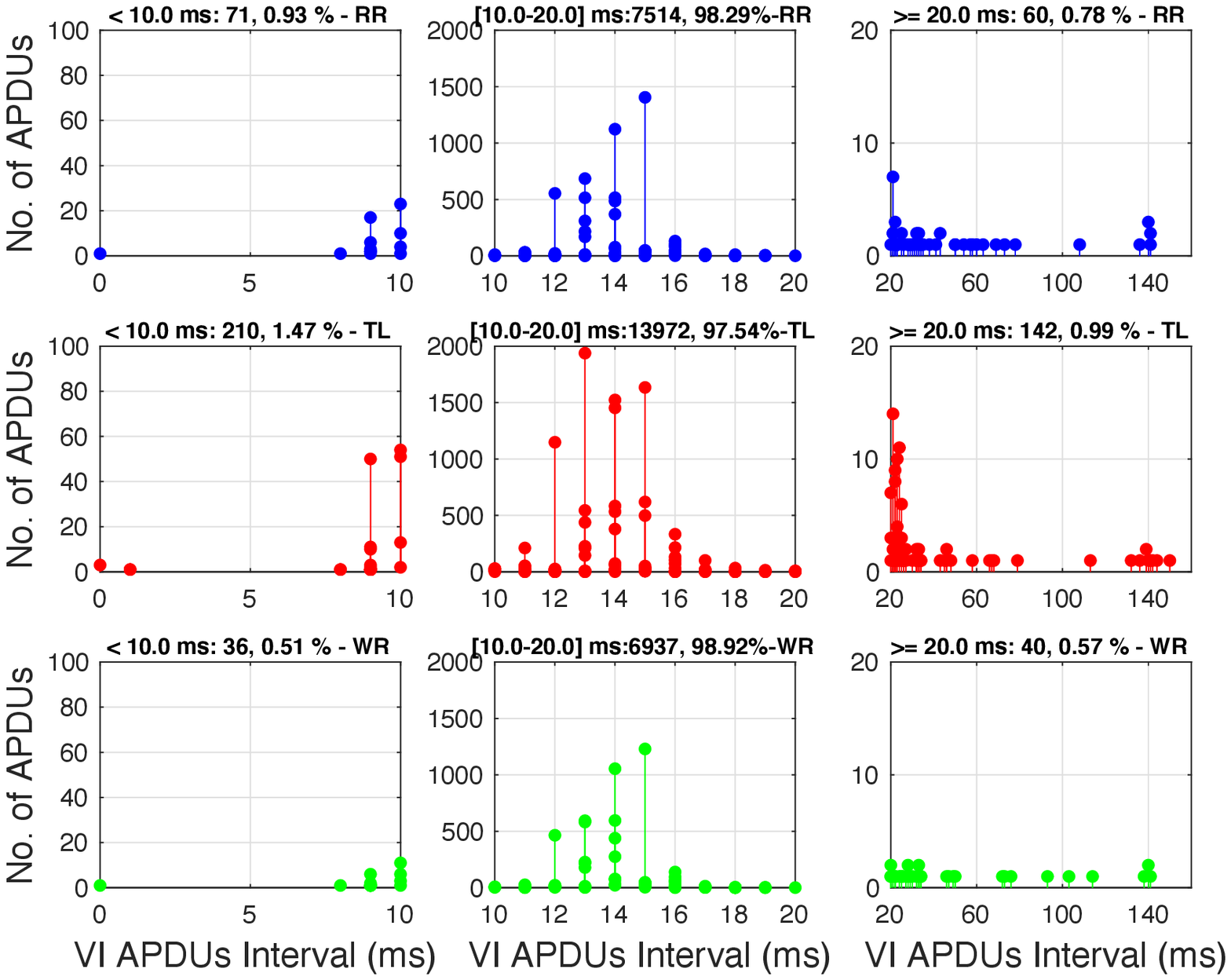} \hspace{-.7em} 
\includegraphics[width=0.505\linewidth]{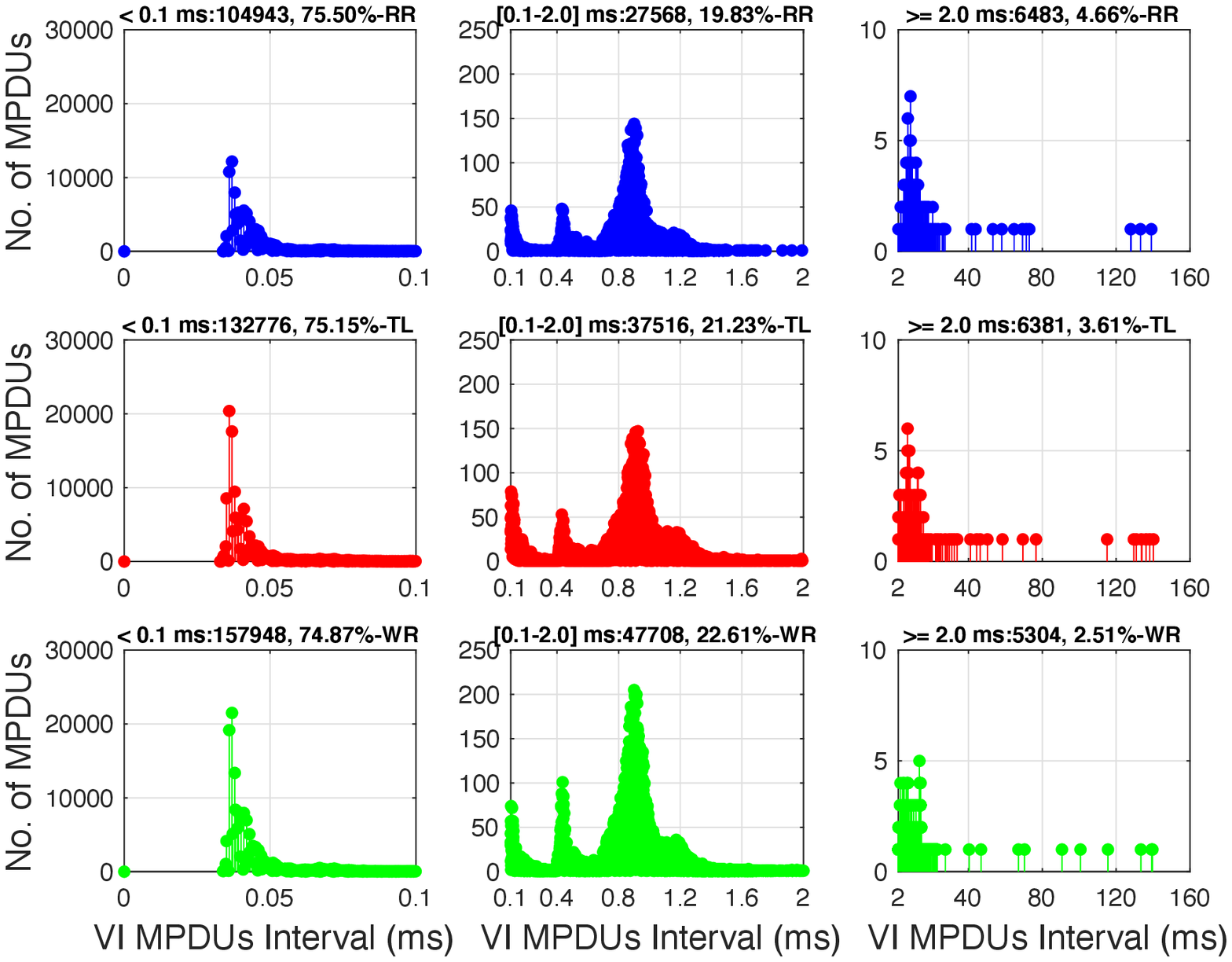}
\caption{Video frame (i.e., APDU) generation interval at the application layer (left) vs. MAC layer (right) for various VR games. The majority of video intervals at the application layer is in the range [12-15] ms. Fragmentation of large video frames into smaller Ethernet frames results in smaller inter-frame interval at the MAC layer (with majority of $\sim$0.04 ms).}
\label{fig:vi_interval}
\end{figure*}

\section{Results and Discussion}
\label{sec:results}

In this section, we compare the statistics of running different VR games with video rendering at the edge-enabled wireless network. Each game is played for 90 seconds starting from the connection establishment between the AVLR client and server (which brings us to the Steam Hall, where we choose a specific game), to load the game and play it as long as the remaining time permits.
Then, we present the statistics of different target bitrates and codecs for the game of Rec Room. 

\begin{figure*}[!htbp]
\centering
\includegraphics[width=0.52\linewidth]{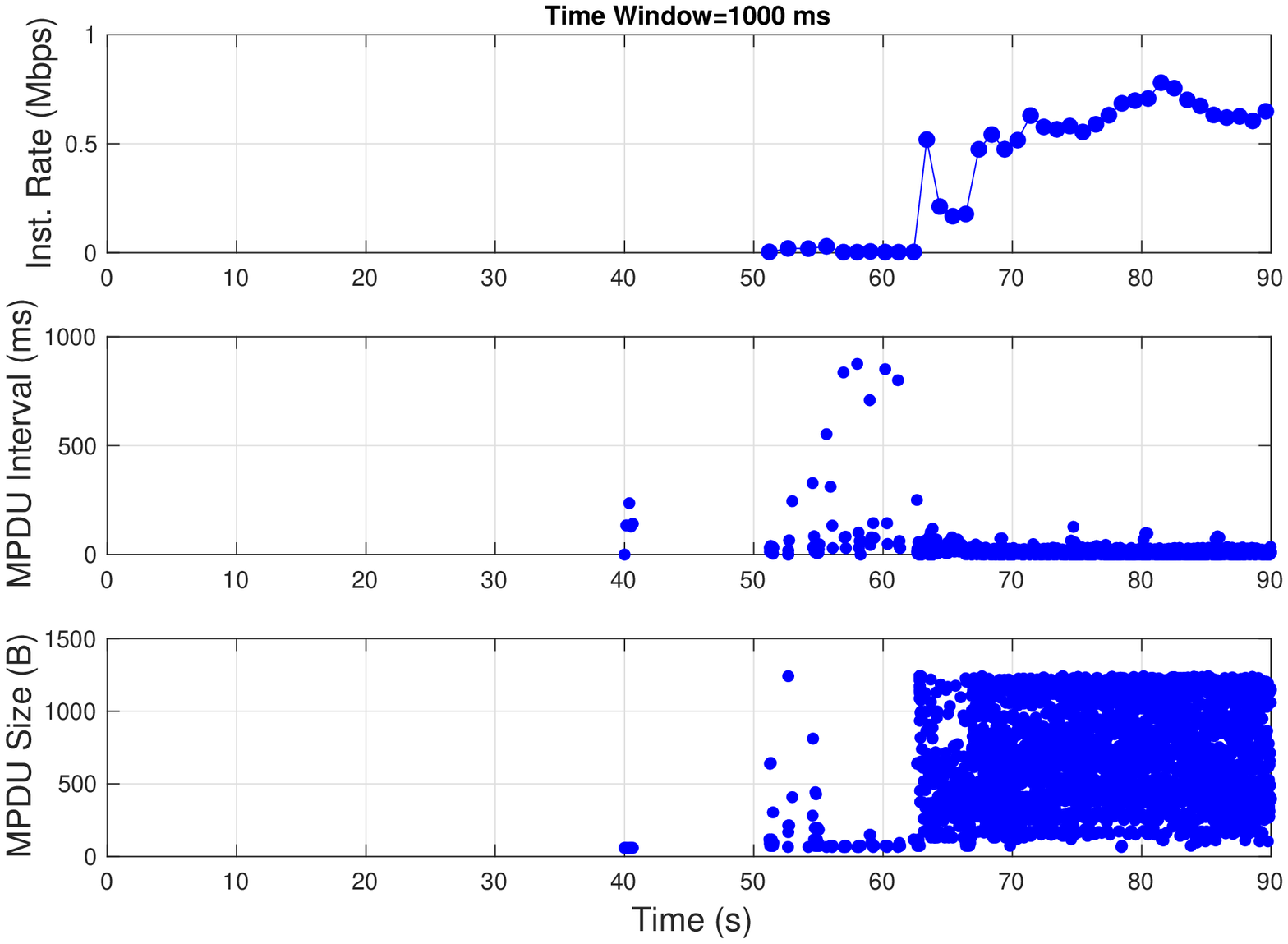} \hspace{-2.8em}
\includegraphics[width=0.52\linewidth]{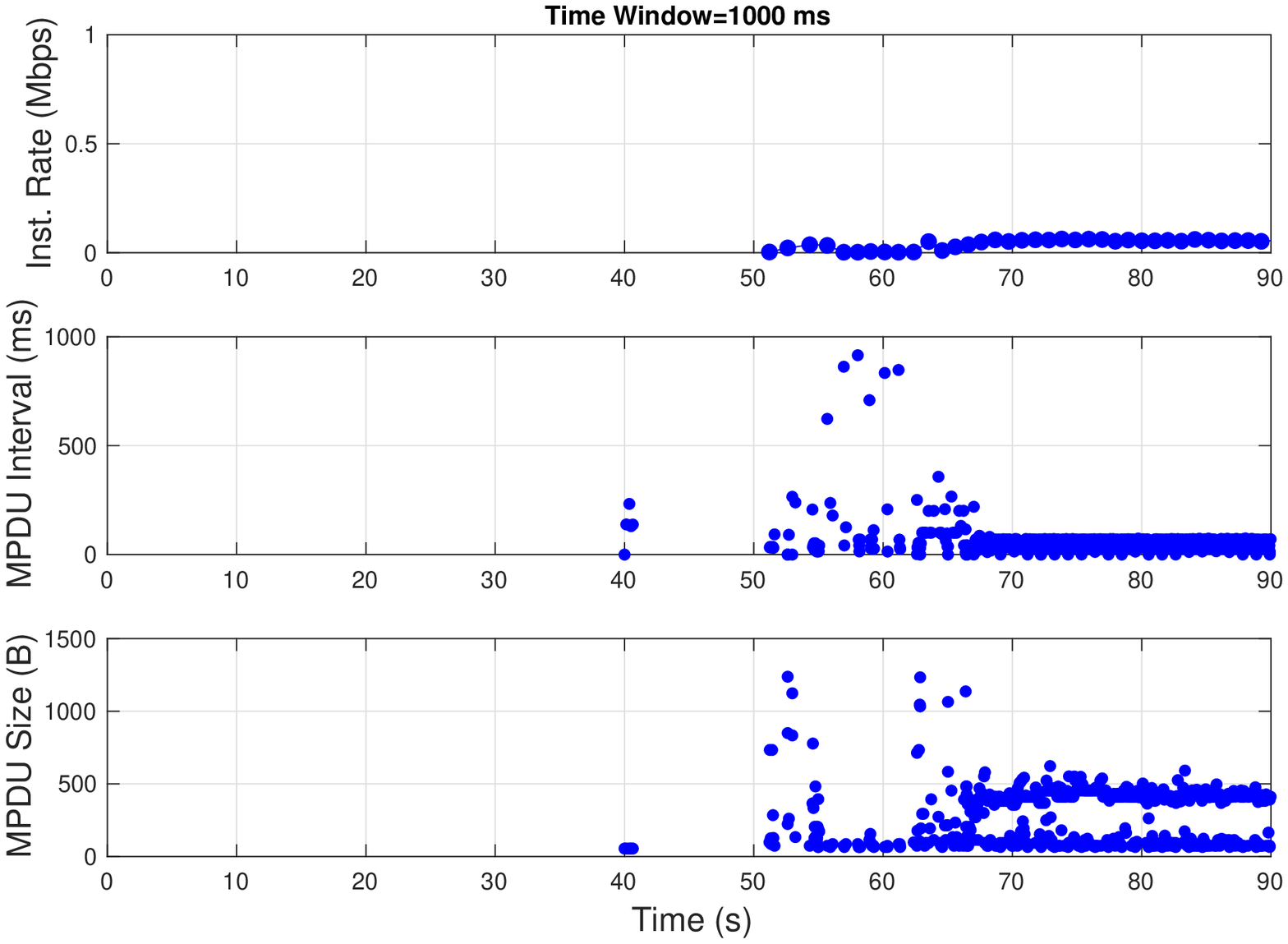}
\caption{MAC layer frame size, interval, and instantaneous rate (with time window = 1 s) of downlink and uplink communications with cloud server for Rec Room. Average rate (from second 40 to second 90) for downlink (left) and uplink (right) is 0.32 Mbps and 0.03 Mbps, respectively.}
\label{fig:size_num_cloud}
\end{figure*}

\begin{figure*}[!htbp]
\centering
\includegraphics[width=0.52\linewidth]{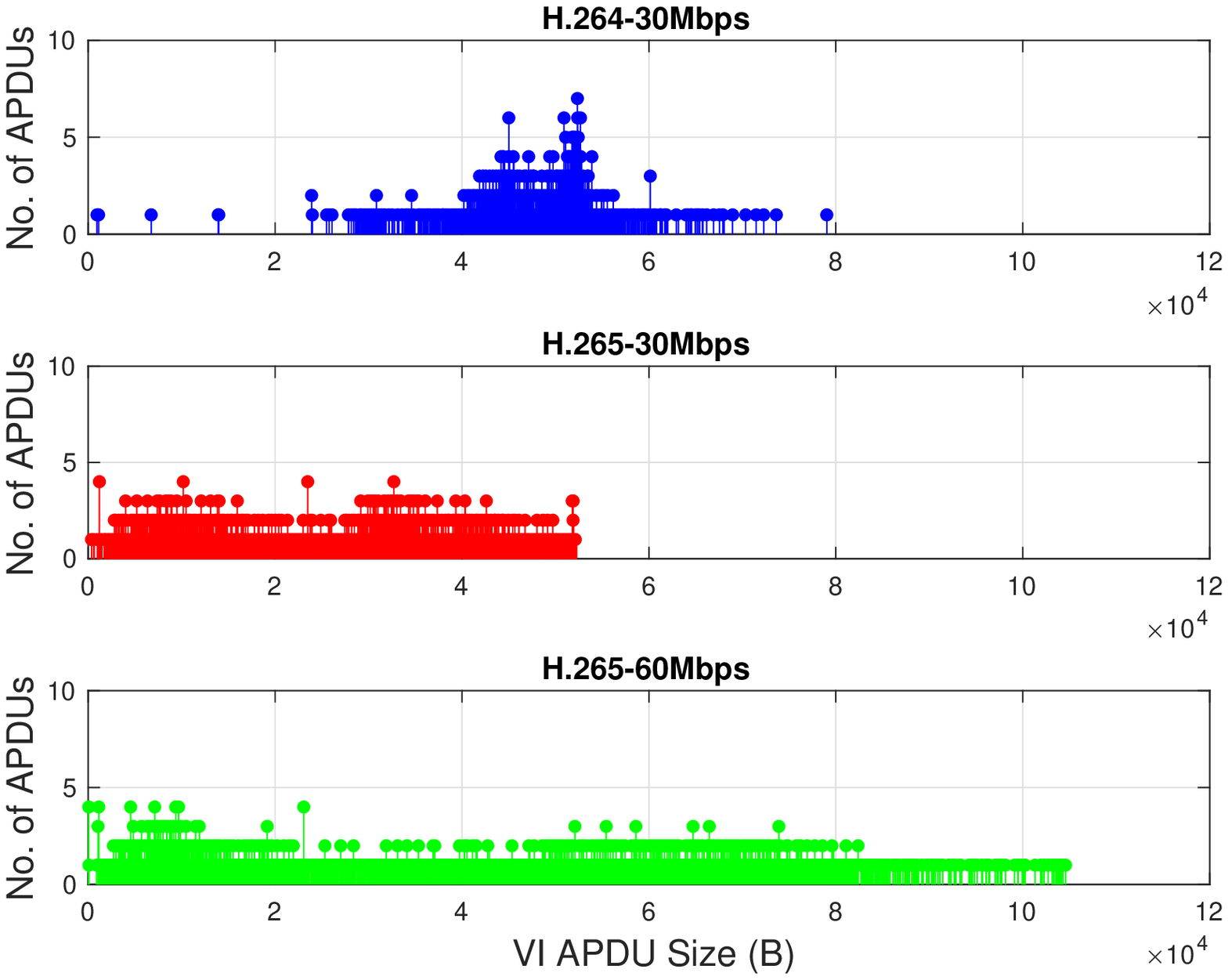} \hspace{-2.8em}
\includegraphics[width=0.52\linewidth]{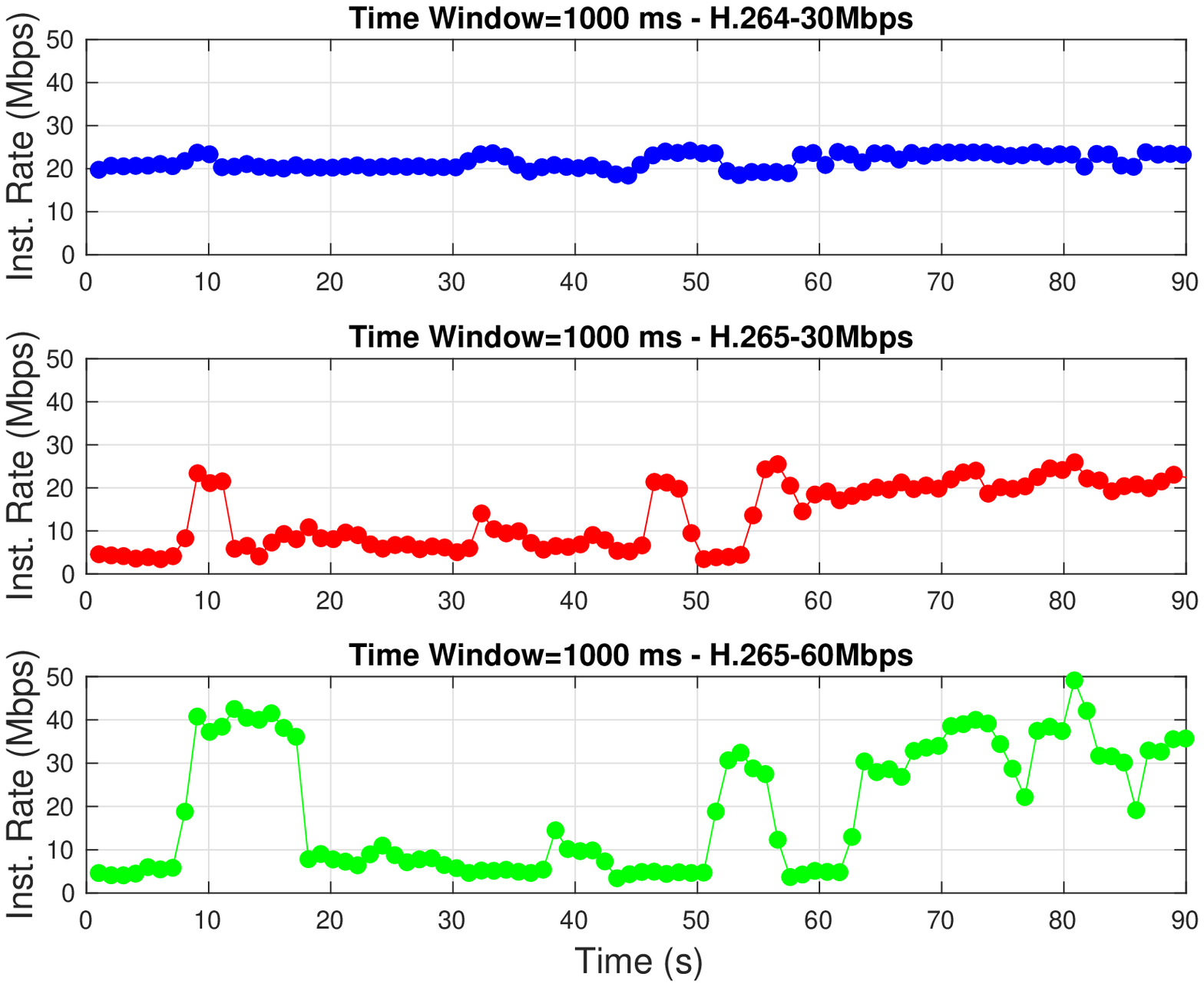}
\caption{Video frame size and instantaneous rate at the application layer for different encoders and target bitrates. The average rate at the Application layer for playing Rec Room with H.264-30Mbps, H.265-30Mbps, and H.265-60Mbps is 21.41, 14.13, and 22.09 Mbps, respectively.}
\label{fig:size_num_encoders}
\end{figure*}

\subsection{Application layer vs. MAC layer}
In this experiment, the target video bitrate is 30 Mbps which reserves a 60 KB buffer size for packets at the client side (i.e., the HMD). Since both H.264 and H.265 use variable bitrate encoding, the actual instantaneous bitrate might be less or more than 30 Mbps depending on the codec type, motion level, frame interval, etc. The frame rate is set to 72 FPS which results in generation of video frames every 13.8 ms.
All voice and video application layer PDUs are sent via UDP protocol. However, owing to the uncrowded network, MAC layer retransmissions and forward error correction (FEC) at the ALVR client (on HMD), the packet loss ratio in our experiment is almost zero. 

Since application video frames constitute more than 90\% of downlink traffic, in this section, we first present the statistics pertaining to generation of the video frames at the application layer and then present the statistics at the MAC layer. Fig. \ref{fig:app_size_num_var} illustrates the quantity and size of generated video frames at the Application layer (i.e., APDUs) for different games with HEVC encoding. With higher graphics demand, larger video frames are generated. For instance, for Rec Room, 47\% of overall video frames are larger than 30 KB. For The Lab and War Robots, 68\% and 74\% of video frames, respectively, are larger than 30 KB. This difference is also shown on the instantaneous rate subfigures where due to the higher graphics demands of the War Robots game, the application layer rate is mostly near 30 Mbps (after second 30 when the game starts). In contrast, for Rec Room, the encoder can reduce the bitrate owing to lower graphics demands.

Fig. \ref{fig:vi_interval} illustrates the interval of video frames generated at the Application (i.e., APDU interval) and MAC layers (i.e., MPDU interval) for different games. Since the frame rate is set to 72 FPS, on average, every 13.8 ms a video frame is generated at the Application layer which gets fragmented into smaller 1442-byte frames at the MAC layer (1400-byte data, 8-byte UDP header, 20-byte IP header, and 14-byte MAC header and trailer). The majority of inter-frame intervals at the MAC layer is in the range [0.03-0.05] ms. 

As mentioned earlier, in MP-VR applications each party also communicates with a cloud server (either directly or through an edge server) to transfer and receive the shared user  information (e.g., voice, updated avatar positions, etc.). Fig. \ref{fig:size_num_cloud} illustrates the cloud communications in DL and UL for Rec Room which starts from time 40 seconds when there is interaction among the players. Depending on the number of parties engaged in a game (or a portion of the game), the amount of DL traffic changes. In this example, the amount of DL traffic is more than 10 times higher than the UL traffic.




\begin{figure*}[!htbp]
\centering
\includegraphics[width=0.52\linewidth]{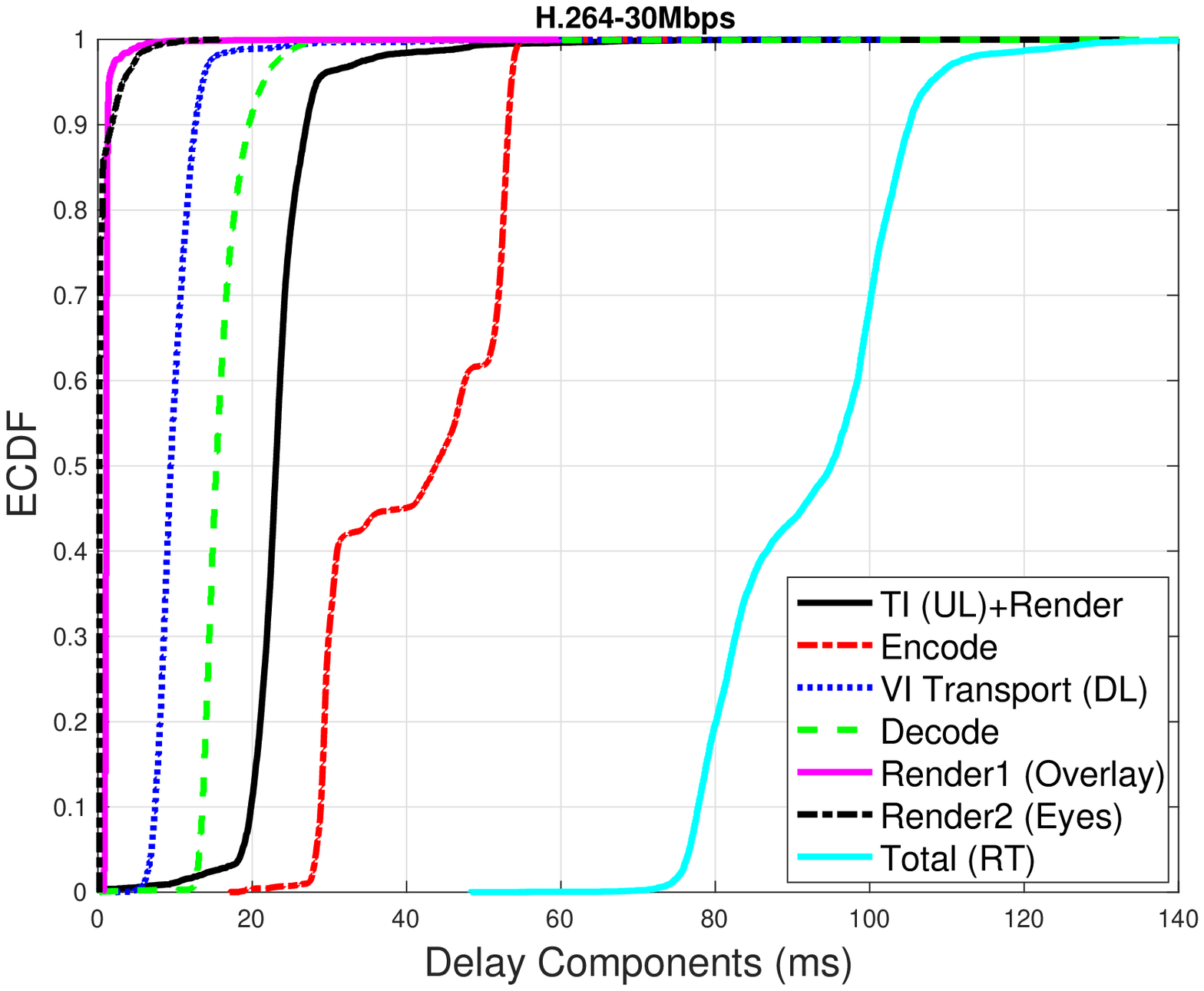} \hspace{-2.8em}
\includegraphics[width=0.52\linewidth]{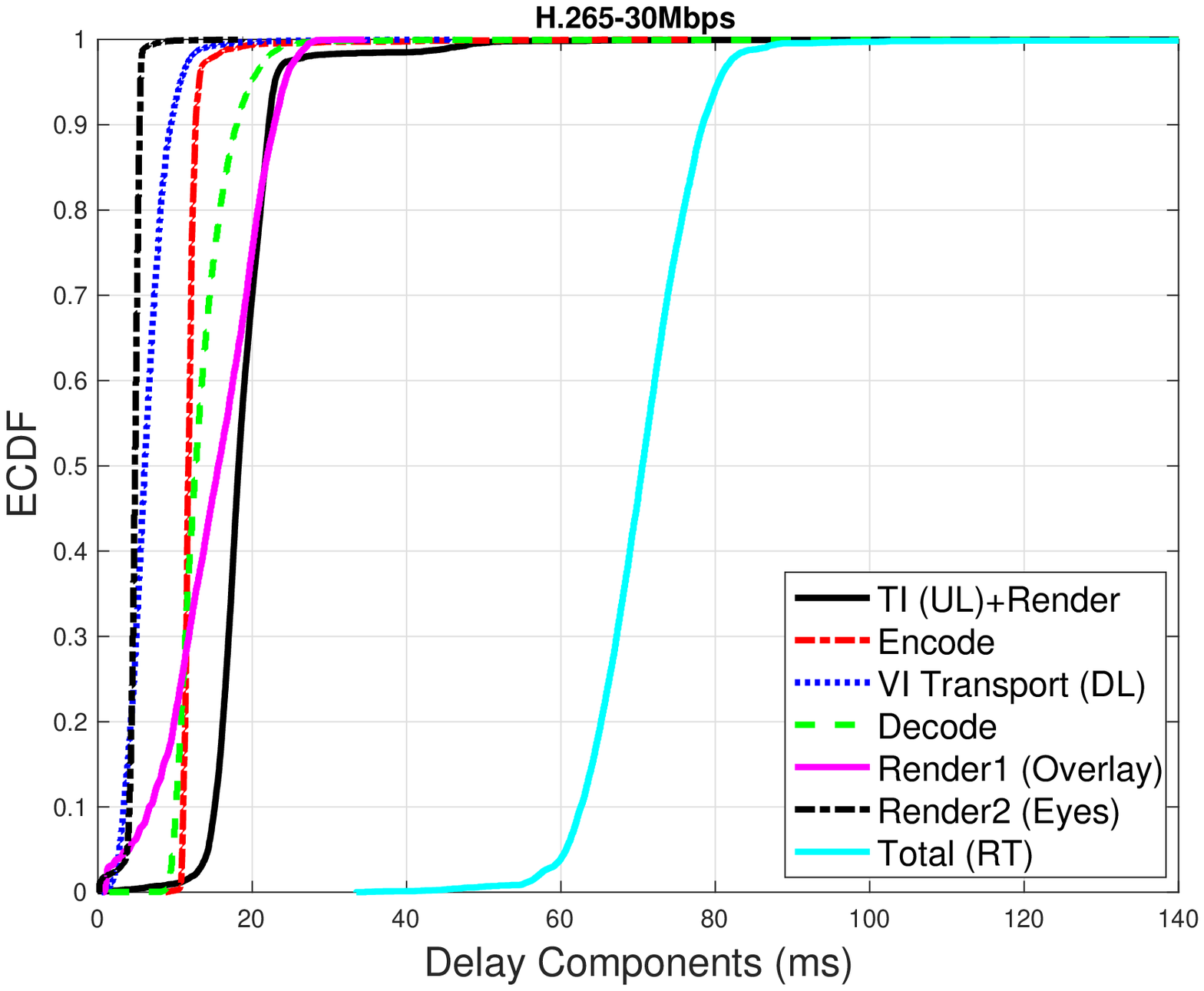}
\caption{Delay components of different encoders with same target bitrate for 90 seconds of playing Rec Room. Although H.265 is superior in the rendering on the edge server and decoding on the HMD, rendering at the HMD is faster with H.264 encoder.}
\label{fig:delay_encoders}
\end{figure*}

\subsection{The impact of different codecs and bitrates on the network}

As video frames of VR applications are large and delay-sensitive, transferring the rendered video frames from the edge server to an HMD incurs a heavy load on the wireless network. Therefore, the remote VR display application should strike a balance among the image quality of video frames, the frame size, and the end-to-end delay. For instance, with a given resolution, higher compression factor produces smaller frame sizes but might exceed the acceptable end-to-end delay. For the AMD graphics cards, ALVR employs Advanced Media Framework (AMF) SDKs\footnote{\url{https://github.com/GPUOpen-LibrariesAndSDKs/AMF}}
for video rendering and encoding. Specifically, ``Ultra-low latency'' usage mode is predefined for video game streaming in which the number of bits per video frame is upper-bounded to the configured target bitrate. 

Fig. \ref{fig:size_num_encoders} depicts a comparison between the video frame sizes and the number of frames for H.264 and H.265 encoders at 30 Mbps target bitrate, and H.265 at 60 Mbps target bitrate. With the same target bitrate of 30 Mbps, the average video frame size generated by H.265 is 24.8 KB which is $\sim$48\% smaller than that of H.264 (i.e., 48 KB). As graphics demand of Rec Room can be met with lower target bitrates, increasing the target bitrate will not increase the clarity of images that constitute the game scenes. However, in high graphics demanding games, higher bitrates can increase the video quality of the games. Therefore, adjustment of target bitrate depending on the VR application can increase the game quality of overall network efficiency.

\begin{table*}[!htbp]
		\centering
		\caption{$\rm{Average}/95^{th}$ percentile of various delay components in milliseconds (ms) for different games and encoders.}
		\label{tab:avg-delay}
		\begin{tabular}{l c c c c c c c}
			\hline
			& {\bfseries TI (UL) + Render} & {\bfseries Encode} & {\bfseries VI Transport} & {\bfseries Decode} & {\bfseries Render1 (Overlay)} & {\bfseries Render2 (Eyes)} & {\bfseries Total (RT)}\\
			\cline{2-8}
			{\bfseries H.265 - 15 Mbps - RR} & $18.57/23.22$ & $12.23/13.78$ & $7.07/11.90$ & $12.40/18.00$ & $14.75/24.20$ & $4.82/5.60$ & $69.85/80.80$
			\\
			{\bfseries H.264 - 30 Mbps - RR} & $23.42/28.35$ & $41.05/53.57$ & $9.99/13.60$ & $16.12/21.50$ & $1.22/1.50$ & $0.67/3.20$ & $92.47/107.90$
			\\
			{\bfseries H.265 - 30 Mbps - RR} & $18.82/22.93$ & $12.01/13.13$ & $6.37/10.70$ & $13.55/19.90$ & $15.17/24.20$ & $4.73/5.50$ & $70.65/80.40$
			\\
			{\bfseries H.265 - 60 Mbps - RR} & $18.17/22.61$ & $11.96/13.04$ & $6.26/10.60$ & $13.38/19.70$ & $14.59/23.90$ & $4.76/5.60$ & $69.11/80.30$
			\\
			{\bfseries H.265 - 30 Mbps - TL} & $18.98/23.38$ & $12.20/14.56$ & $7.75/11.70$ & $12.20/17.58$ & $16.72/24.98$ & $4.83/5.60$ & $72.67/81.20$
			\\
			{\bfseries H.265 - 30 Mbps - WR} & $19.55/23.72$ & $11.88/12.78$ & $8.98/13.60$ & $11.88/16.70$ & $14.02/24.10$ & $4.76/5.60$ & $71.07/81.80$
			\\
			\hline
		\end{tabular}
\end{table*}

As mentioned ealier, to generate a video frame, the game engine requires the updated position and velocity of each engaged player.
As an example, for Rec Room, Fig. \ref{fig:delay_encoders} illustrates the CDF statistics of delay components for H.264 and H.265 at target bitrate of 30 Mbps. Total delay is round trip which is computed from the time a tracking information APDU is sent to the edge server until the time that the video frame (scene) is displayed at the HMD screen(s). Since the frequency of transmitted tracking information packets might be higher than the video frame rate, the game engine only considers the most updated tracking information in generating the next video frame.
Therefore, rendering delay at the edge server is coupled with tracking information transmission delay. Although H.265 is superior to H.264 in most of the delay compartments, rendering delays on the HMD for H.264 are smaller than H.265. In addition, unlike H.265, H.264 is unable to encode the video frames within the bounds of frame rate (on average encoding with H.264 takes longer than $1/72=13.88$ ms). Table \ref{tab:avg-delay} contains the average and $95^{th}$ percentile statistics of delay components for different games and target bitrates. Running Rec Room with the target bitrate of 15 Mbps results in reduced quality of scenes. However, for Rec Room, there is no noticeable advantage in increasing the target bitrate to 60 Mbps compared to 30 Mbps. Therefore, adjustment of the target bitrate should be based on the graphics demands of the games that can be automated to strike a balance between the gaming experience and the network load.

 
\subsection{Discussion}
As we discussed earlier, transmission of tracking information frames from the HMD to the edge server with high frequency can increase the overall delay, especially in densely-deployed or crowded wireless networks. Instead, the frequency of tracking information can be based on the instantaneous VR experience. Even with fixed tracking information intervals, unlike IEEE 802.11ac, IEEE 802.11ax can obtain the tracking interval of multiple STAs simultaneously and reduce the channel contention which requires a scheduling mechanism that gradually learns and synchronizes the tracking information intervals of different HMDs.

With the same frame rate and target bitrate, compared to H.264, H.265 encoder ends up using less rate and generating smaller video frames (on average 49\% smaller), and is able to meet the configured frame rate. Although increasing the target bitrate results in sharper and more clear video frames, it also generates larger video frames that incur challenges on the Wi-Fi network. Therefore, adjustment of the target bitrate should be based on the resolution requirements of the games and can be automated to strike a balance between the gaming experience and the network load.



In our experiment, the packet loss ratio is almost zero because of the uncrowded network, high SNR at the HMD, MAC layer retransmissions, and forward error correction (FEC) at the HMD. In densely-deployed or crowded wireless networks, for an immersive VR experience, higher throughput, lower contention, and reliability are required. IEEE 802.11ax offers the simultaneous transmission of downlink or uplink frames via OFDMA. Therefore, new scheduling mechanisms are required to satisfy the QoS requirements of the VR applications while not starving other high priority and delay-sensitive applications (i.e., regular video and voice). Since the QoS requirements of VR applications is more stringent than that of regular video applications, new access categories might be needed to prioritize VR traffic over regular interactive video applications.


\section{Conclusion and Future research directions}
\label{sec:conclusion}

VR is a promising technology that facilitates immersive user experience in many fields. Due to the heavy 3D computational requirement of VR applications, rendering can be offloaded to an edge server which reduces the energy consumption and production cost of the VR HMD. In this paper, we theoretically computed the throughput requirements of an eye-like experience as well as a popular VR HMD (i.e., Oculus Quest). We then presented the realistic network statistics of playing different VR games in an edge-enabled wireless network. These network statistics can be used in other trace-driven simulations to develop new architectures, protocols, access mechanisms and scheduling algorithms.

Future research can focus on the optimization of (1) tracking information intervals (e.g., frequency adjustment based on the instantaneous VR experience and frame rate), and (2) target bitrate based on different games and instantaneous game situations. In the MAC layer, we plan to investigate a new MAC layer access category for VR traffic and new scheduling mechanisms so as to satisfy the QoS requirements of VR applications while not starving other traffic types. 

\section*{Acknowledgement}
This study has been supported by a gift grant from Cisco Systems, Inc.

\bibliographystyle{IEEEtran}
\bibliography{references.bib}

\end{document}